\documentclass[twoside,a4paper,11pt]{sea10}
\usepackage{graphicx}
\usepackage[dvipsnames]{color}
\usepackage{hyperref}
\usepackage{amssymb}
\usepackage{movie15}
\newcommand{\de} {{\rm d}}

\begin{document}
\pagenumbering{arabic}
\pagestyle{myheadings}
\thispagestyle{empty}
\vspace*{0.2cm}

\begin{flushleft}
{\bf {\LARGE Magnetic fields in Neutron Stars}\\
\vspace*{1cm}
Daniele Vigan\`o$^{1,2}$, Jose A. Pons$^2$, Juan A. Miralles$^2$, Nanda Rea$^3$}\\
\vspace*{0.5cm}
$^{1}$ Institute of Space Sciences (CSIC--IEEC), Campus UAB, Faculty of Science, Torre C5-parell, E-08193 Barcelona, Spain\\
$^{2}$ Departament de F\'{\i}sica Aplicada, Universitat d'Alacant, Ap. Correus 99, 03080 Alacant, Spain\\
$^{3}$ Anton Pannekoek Institute, University of Amsterdam, Postbus 94249, 1090GE Amsterdam, The Netherlands.\\
\end{flushleft}

\markboth{Magnetic fields in Neutron Stars}{Daniele Vigan\`o}
\thispagestyle{empty}
\vspace*{0.4cm}
\begin{minipage}[l]{0.09\textwidth}
\ 
\end{minipage}
\begin{minipage}[r]{0.9\textwidth}
\vspace{1cm}
\section*{Abstract}{\small
Isolated neutron stars show a diversity in timing and spectral properties, which has historically led to a classification in different sub-classes. The magnetic field plays a key role in many aspects of the neutron star phenomenology: it regulates the braking torque responsible for their timing properties and, for magnetars, it provides the energy budget for the outburst activity and high quiescent luminosities (usually well above the rotational energy budget). We aim at unifying this observational variety by linking the results of the state-of-the-art 2D 
magneto-thermal simulations with observational data. The comparison between theory and observations allows to place two strong constraints on the physical properties of the inner crust. First, strong electrical currents must circulate in the crust, rather than in the star core. Second, the innermost part of the crust must be highly resistive, which is in principle in agreement with the presence of a novel phase of matter so-called nuclear pasta phase.
\normalsize}
\end{minipage}

\section{Introduction \label{intro}}

Isolated neutron stars (NSs) are the strongest magnets in the universe. The magnetic fields inside and outside NSs are responsible for the peculiar properties and phenomena observed in the radio, X-ray and $\gamma$-ray bands. Timing properties (spin period and its time derivative) are regulated by the magnetospheric electro-magnetic torque that spins down the star. By measuring these two quantities, we can infer the strength of the surface magnetic field at the pole, $B_p$, which, in turn, depends on the internal magnetic field. The latter is sustained by electrical currents that can circulate in the highly conducting core and in the solid crust. Magnetic fields can be so strong that they can power the thermal emission via Joule dissipation of the electrical currents that circulate in the NS crust. Therefore, the X-ray luminosity of the most magnetic stars can be significantly higher than that of normal pulsars. Furthermore, the presence of strong magnetic fields directly affects the microphysical processes that govern the thermal evolution of the crust, as the temperature in turn modifies the magnetic field evolution.

This coupling between thermal and magnetic evolution has to be included to study the cooling of NSs beyond the classical 1D approach, developed short after the first hints of X-ray emission from the surface of NSs \cite{morton64,chiu64,tsuruta65}. Such models take into account the necessary microphysics in order to simulate the radial transfer of internal heat to the surface. More recent works \cite{aguilera08b,pons07b,pons09} paved the way for a fully coupled magneto-thermal evolution. In the PhD thesis \cite{viganothesis}, we have produced the first results coming from 2D simulations that properly include the important Hall term in the induction equation describing the magnetic field evolution. This allows us to follow the long-term evolution of magnetized NSs and explain the varied phenomenology in terms of the different initial magnetic field and age.

In \S\ref{sec:observations} we present the observational X-ray sample of thermally emitting isolated NSs. In \S\ref{sec:cooling} we review the theory of magneto-thermal evolution of isolated NSs. In \S\ref{sec:results} we show the results of the simulations, and the implications for the unification of the sub-classes of isolated NSs. We draw the conclusions in \S\ref{sec:conclusions}.

\section{Data on cooling NSs.}\label{sec:observations}

To confront theoretical cooling models of NSs with the observational data, we need to know simultaneously the age and the thermal luminosity. The number of sources for which both measures are available is limited, and in most cases subject to large uncertainties. 
All the timing and spectral properties (with abundant references) can be found in our website.\footnote{\href{http://www.neutronstarcooling.info}{http://www.neutronstarcooling.info}}

From the timing properties one can infer the surface magnetic field at the pole, $B_p$, by equating radiative and rotational energy losses:
\begin{equation}\label{eq:spindown_forcefree}
I\Omega \dot{\Omega} =  \frac{B_p^2R_\star^6\Omega^4}{6 c^3}f_\chi~.
\end{equation}
where $I$ is the moment of inertia, $R_\star$ is the NS radius, $\Omega=2\pi/P$ is the angular frequency. Differences in the radiation mechanism are included in the factor of order unity $f_\chi$, which is $f_\chi=\sin\chi$ for a spinning dipole in vacuum or $f_\chi=1.5(1+\sin^2\chi)$ for force-free magnetospheres \cite{li12a}. Here $\chi$ is the angle between the rotation and magnetic axis.

In terms of the spin period and its derivative, we can write in compact form:
\begin{equation}\label{eq:ppdot_spindown} 
  P \dot{P} = K B_p^2~,
\end{equation}
with
\begin{equation}\label{eq:k_spindown}
  K=f_\chi\frac{2\pi^2}{3}\frac{R_\star^6}{Ic^3}=  2.44\times 10^{-40}~ f_\chi \frac{R_6^6}{I_{45}} \mbox{~s G}^{-2}~,
\end{equation}
where $R_6=R_\star/10^6~$cm and  $I_{45}=I/10^{45}$ g cm$^2$. In the literature, the fiducial values $I_{45}=1$, $R_6=1$,
and $f_\chi=1$ are the most common choice (but not the most accurate) used to infer the value of $B_p$.

The characteristic age $\tau_c=P/2 \dot{P}$ can be used as an approximation to the real age. Usually, for middle-aged and old objects, $\tau_c$ is found to be larger than the real age, when the latter has been obtained by other methods. When the object is located in a SNR, a kinematic age can also be inferred by studying the expansion of the nebula. Alternatively, the proper motion, combined with an association to a birth place, can also give an estimate of the age \cite{tetzlaff11,tendulkar12}. We have collected the most updated and/or reliable available information on timing properties and kinematic ages from the literature, the ATNF catalogue\footnote{\tt http://www.atnf.csiro.au/people/pulsar/psrcat/} \cite{manchester05}, and the McGill online magnetar catalogue\footnote{\tt http://www.physics.mcgill.ca/$\sim$pulsar/magnetar/main.html}\cite{olausen14}.

Luminosities and temperatures can be inferred by spectral analysis, but it is usually difficult to determine them accurately. The luminosity is always subject to the uncertainty in the distance measurement, while the inferred effective temperature depends on the choice of the emission model (blackbody versus atmosphere models, composition, condensed surface, etc.), which is subject to large theoretical uncertainties in the case of strong magnetic fields.  Often more than one model can fit equally well the data, with inferred effective temperatures differing up to a factor of two. Photoelectric absorption from interstellar medium further constitutes a source of error in temperature measurements. Overall, the luminosity constitutes a better choice to compare data and theoretical models.

We have homogeneously re-analysed all the data of our sample, extracted directly from the best available observations from {\em Chandra/EPIC-pn} or {\em XMM--Newton/ACIS-S}. Standard data screening criteria and analyses procedures are applied. For the spectral analysis we used the {\tt XSPEC} package (version 12.4) for all fittings, and the {\tt phabs} photoelectric absorption model with the \cite{anders89} abundances, and the \cite{balucinska92} photoelectric cross-sections. We usually restricted our spectral modelling to the 0.3--10~keV energy band, unless the source was such that a smaller energy range was needed for the spectral analysis. 

The sample of selected sources includes all NSs which thermal emission is clearly detected (in their quiescent state), and information about age and distance are also available. We exclude those objects for which the spectrum was fitted equally well without the addition of a thermal component. The final sample includes:

$\bullet$ 4 Central Compact Objects (CCOs), which are a class of young, radio-quiet, X-ray sources \cite{gotthelf13} located close to the center of $\sim$ kyr old supernova remnants (SNRs). They show very stable, thermal-like spectra, with hints of temperature anisotropies, in the form of large pulsed fraction, or small emitting regions characterized by hot ($0.2$--$0.4$ keV) blackbody components. The period and period derivative are known for only three cases, and their spin period is basically the same as the natal one. For this reason, their characteristic ages are much longer than their real ages. The dipolar component of the external magnetic field is low, $B_p\sim 10^{10}$--$10^{11}$~G.\\

$\bullet$ 12 rotationally powered-pulsars (RPPs), including 4 high-B pulsars \cite{ng11}. Their X-ray spectra are described by thermal plus non-thermal components. In particular, the three Musketeers (Geminga, PSR~B0656 and PSR~B1055; \cite{deluca05}) are well fitted by a double blackbody plus a power law. An atmosphere model (plus a power law) fits Vela better than a blackbody plus power law, and it is compatible with emission from the entire surface. In general, the RPP sub-class consists of than 2000 pulsars detected in radio, and more than one hundred in $X$-rays and/or $\gamma$-rays, with spin periods in the range 1.3 ms--8.5 s, with inferred magnetic fields ranging from $\sim 10^8$ G to $\sim 10^{14}$ G.\\

$\bullet$ 7 radio-quiet X-ray isolated NSs (XINSs, \cite{haberl07,turolla09,kaplan09b}). They are relatively old ($\sim 10^5$ yr), nearby NSs, with the cleanest detected thermal emission and a relatively large magnetic field and long periods (3-11 s). Their proximity ($d< 1$ kpc) makes them the most promising candidates to study NS thermal spectra, which are well fit by a single blackbody model, with absorption features in a few cases.\\

$\bullet$ 17 of the $\sim 25$ magnetar candidates \cite{olausen14,thompson95}: those having good spectra in quiescence, i.e., not during transient outbursts often shown by these sources \cite{rea11}.  They rotate relatively slow, with spin periods $P\sim 2-12$ s, and $B_p\sim 10^{14}-10^{15}$ G. For the spectral fit, we have also used a resonant Compton scattering model ({\tt RCS}: \cite{rea08,lyutikov06}), adding a power law component when needed.

\section{Cooling}\label{sec:cooling}

In the presence of magnetic fields, the heat and electrical conduction is anisotropic, and one has account for the magnetic energy dissipated by the Joule effect \cite{haensel90,miralles98,page00,tauris01,pons09}. Therefore, the energy balance equation reads
\begin{equation}\label{eq:heat_balance}
c_v e^\nu\frac{\partial T}{\partial t} + \vec{\nabla}\cdot(e^{2\nu}\vec{F}) = e^{2\nu}({\cal Q}_j - {\cal Q}_\nu)~,
\end{equation}
where $c_v$ is the heat capacity per unit volume and $\vec{F}$ is the thermal flux, which
in the diffusion limit is given by
\begin{equation}\label{eq:flux_diffusion} 
\vec{F} = -e^{-\nu} \hat{\kappa} \cdot \vec{\nabla} (e^\nu T) ~,
\end{equation}
where $\hat \kappa$ is the thermal conductivity tensor.
${\cal Q}_\nu$ are the neutrino emission losses per unit volume and time, and
\begin{equation}\label{eq:def_joule}
 {\cal Q}_j = \frac{J^2}{\sigma}
\end{equation}
is the Joule dissipation rate of the electrical currents, defined as $\vec{J}= e^{-\nu} (c/4\pi) (\vec{\nabla}\times(e^\nu\vec{B}))$.  The electrical conductivity $\sigma$ is dominated by the electronic transport, and depends on different scattering processes which define the electron relaxation time $\tau_e$: $\sigma=e^2n_e\tau_e/m^\star_e$, where $m^\star_e$ and $n_e$ are the effective mass and the electron density. The relaxation time strongly depends on the temperature and the impurity concentration within the crust. Since the electron density varies over about four orders of magnitude in the crust and the temperature decreases by about two to three orders of magnitude in a pulsar's lifetime, the electric conductivity may vary both in space and time by many orders of magnitude.

Above, we have considered the standard, spherically symmetric, static metric
\begin{equation}\label{eq:metric}
 \de s^2 = - c^2 e^{2\nu(r)}\de t^2 + e^{2\lambda(r)}\de r^2 + r^2\de\theta^2 + r^2\sin^2\de\varphi^2~,
\end{equation}
where $e^{\lambda(r)} = (1 - 2Gm(r)/c^2r)^{-1/2}$ is the space curvature factor, $m(r)$ is the enclosed gravitational mass within radius $r$, and $e^\nu(r)$ is the lapse function that accounts for redshift corrections.

The thermal evolution is coupled to the magnetic field evolution. In the crust, ions form a Coulomb lattice, while electrons are relativistic, degenerate and can almost freely flow. This limit is known as electron MHD (EMHD), and it is apt to describe the solid crust, where electrons are the only moving particles and carry all the current, $\vec{J}=-en_e\vec{v}_e$. Then, the evolution of the system is governed by the Hall induction equation which has the form \cite{pons09}:
\begin{equation}\label{eq:induction}
 \frac{\partial \vec{B}}{\partial t} =  -\vec{\nabla}\times \left[\frac{c^2}{4\pi \sigma} \vec{\nabla}\times(e^\nu\vec{B}) + \frac{c}{4\pi e n_e}  (\vec{\nabla}\times(e^\nu\vec{B}))\times\vec{B}\right]~, \label{eq:magnetothermal_b}
\end{equation}
The first (parabolic) term on the right hand side accounts for Ohmic dissipation, while the second term arises because of the Hall effect. 
When the Hall term dominates, the induction equation acquires a hyperbolic character. In this case, electric currents are squeezed in a smaller volume and move through the crust, creating small scale structures and allowing the energy interchange between poloidal and toroidal components of the magnetic field.
%

To solve the coupled heat balance and induction equations, we first need to adopt a background NS model, by means of an equation of state, as well as to calculate all microphysical inputs: specific heat, thermal and electrical conductivities and neutrino emissivities. These ingredients depend in general on the local values of temperature, density, composition and magnetic field strength. Among the parameters regulating the magneto-thermal evolution, we mention the impurity parameter $Q_{\rm imp}$, an effective value which describes how crystalline/amorphous is the lattice. Large values of $Q_{\rm imp}$ effectively describe a lattice composed by different species of ions (impure), or disposed in an amorphous state, e.g., due to the presence of the pasta phase \cite{pons13}. In such cases, the conductivity is lower than for low $Q_{\rm imp}$, which implies shorter timescales of the magnetic field dissipation. Other microphysical/NS parameters, like the mass, the equation of state, the superfluid gap model, are more relevant for models with $B\lesssim 10^{14}$ G (see \cite{vigano13,aguilera08b} for details).

\section{Results.}\label{sec:results}

\subsection{Cooling of weakly magnetized NSs.}\label{sec:cooling_b0}

\begin{figure}
 \centering
\includegraphics[width=.47\textwidth]{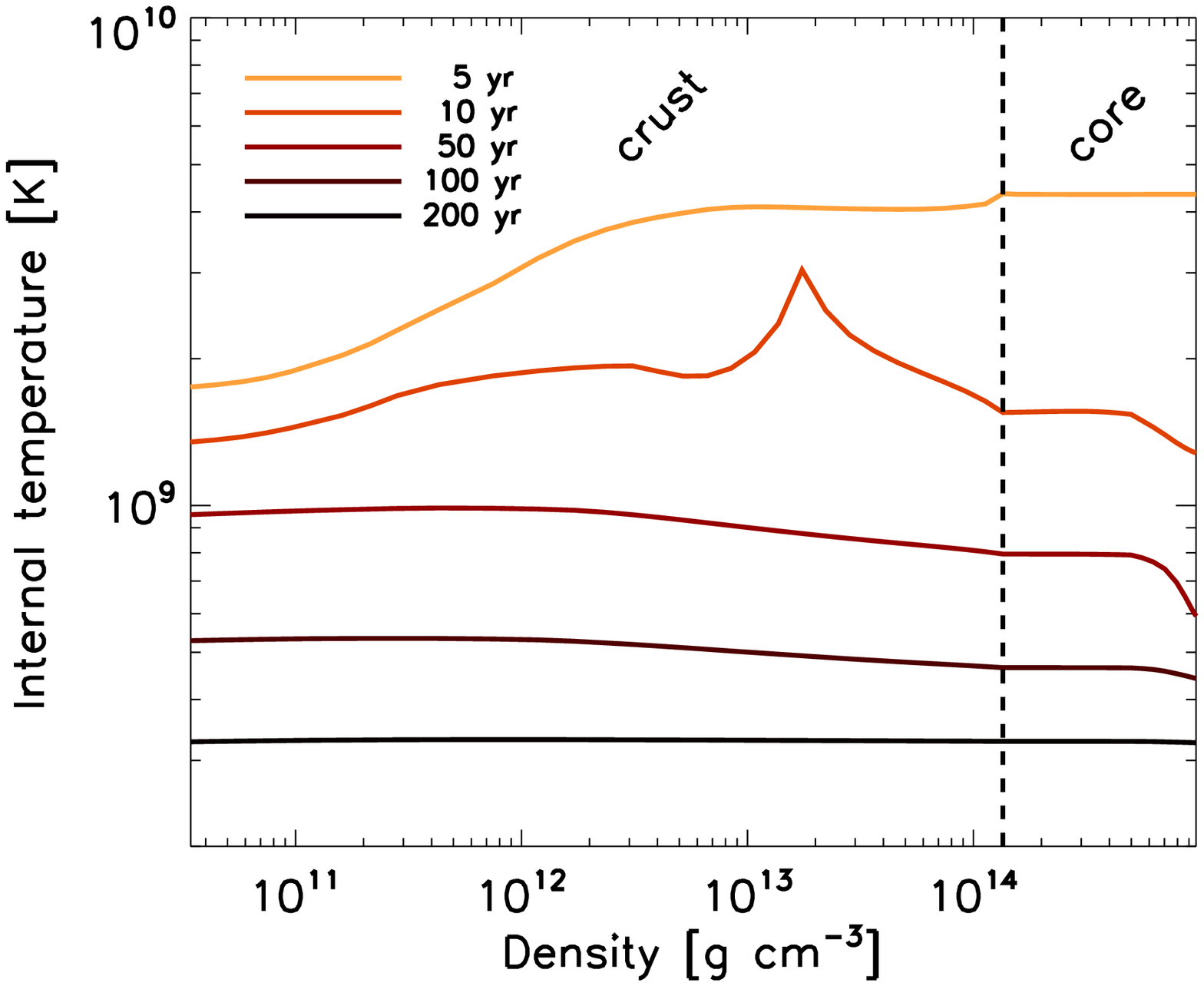}
\includegraphics[width=.47\textwidth]{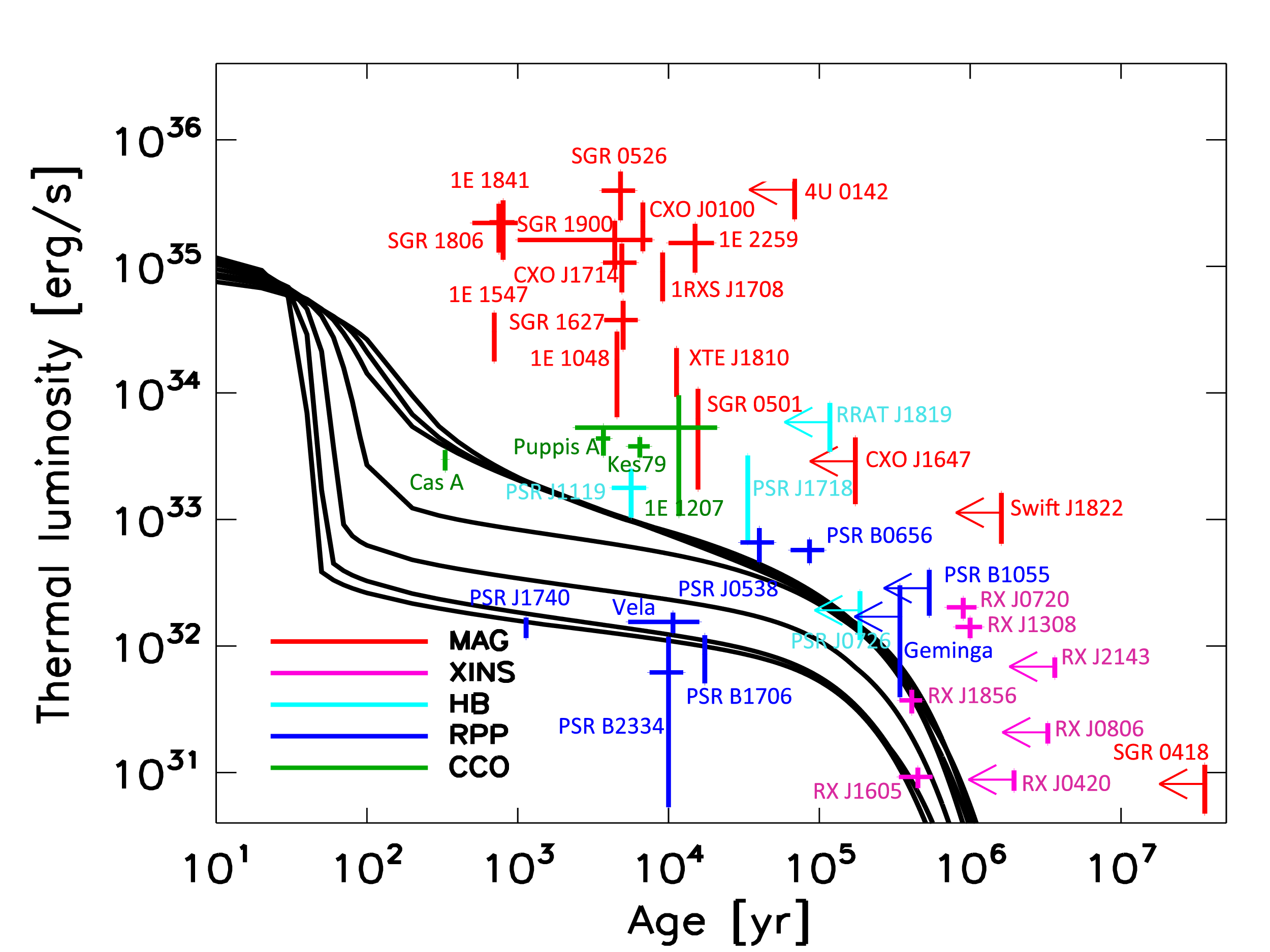}
\caption{Left: evolution of internal temperature in the early stages ($t=5,10,50,100$ and $200$ yr), starting with $T_{init}=10^{10}~$K, for a non-magnetized NS with $M=1.4~M_\odot$. Right: luminosity versus age for non-magnetized NS models.  Arrows label sources for which $\tau_c\gtrsim 50$ kyr, and no kinematic age is available, so that the real age is expected to be shorter. An uncertainty of $50\%$ has been arbitrarily taken for the kinematic age when error estimates have not been found in the literature. Taken from \cite{viganothesis}.}
 \label{fig:tint_b0}
\end{figure}

In the absence of magnetic field, the cooling models simulate 1D radial transfer of the residual heat stored in the NS at birth. The evolution of the internal temperature in the early stages is shown in the left panel of Fig.~\ref{fig:tint_b0}. The temperature, $T_{init} \gtrsim 10^{10}$ K at the beginning, drops below $10^9$ K after a few decades, because of the very efficient neutrino emission in the inner core. The crust, with a larger thermal relaxation time-scale (a few years), follows the core cooling. After a few centuries, the star becomes isothermal, except in the envelope (not shown, $\rho \lesssim 10^{10}$ g~cm$^{-3}$), where a strong temperature gradient is present at any age. Meanwhile, the temperature in the region $\rho\sim 10^{11}$--$10^{14}$ g~cm$^{-3}$ gradually drops below the melting temperature and the hot 
and high-$Z$ ion-rich plasma solidifies to form the crust.

In the right panel, we show the cooling curves for masses ranging between 1.10 and 1.76 $M_\odot$, and for models with iron envelopes. After $\approx 100$ yr, low mass stars are hotter and brighter than high mass stars. For the high-mass family, $M > 1.4~M_\odot$, the activated direct URCA processes result in fast cooling before $\sim 100$ yr. Within the low-mass family, cooling curves are similar at all ages. The small differences at $t\sim 10^2$--$10^3$ yr depend on the superfluid gap and the equation of state employed. After the transition to a superfluid core, cooling curves for low mass NSs tend to converge again, following the same curve independently of the mass. We also show the luminosities extracted from observational data. Sources with estimates of the kinematic age are shown with the associated error bar on the age. We indicate with arrows pointing towards the left the sources with $\tau_c\gtrsim 50$ kyr and without a measure of the kinematic age, since magnetic field dissipation implies that the characteristic age overestimates the real age.

Objects with larger inferred magnetic field (magnetars in particular) are systematically hotter than what theoretical non-magnetized cooling curves predict. This provides a strong evidence in favor of the scenario in which magnetic field decay powers their emission. Even considering the (more than likely) overestimate of the NS age by using its characteristic age, which can reconcile some of the objects with standard cooling curves, it is clear that most magnetars and some high B NSs cannot be explained.

\subsection{Magnetic field evolution.}\label{sec:magnetic}

\begin{figure}[t]
\centering
\includegraphics[width=.32\textwidth]{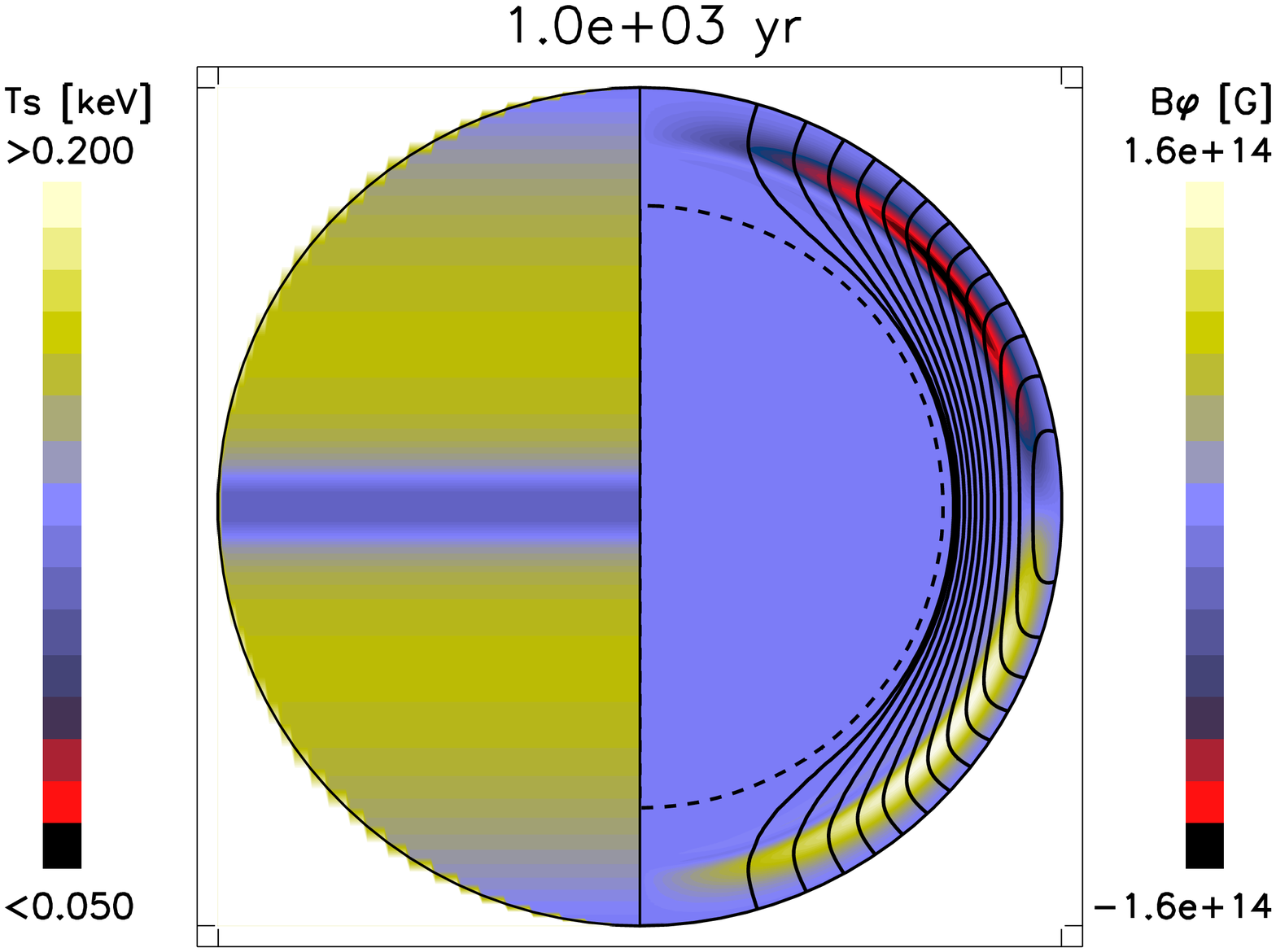}
\includegraphics[width=.32\textwidth]{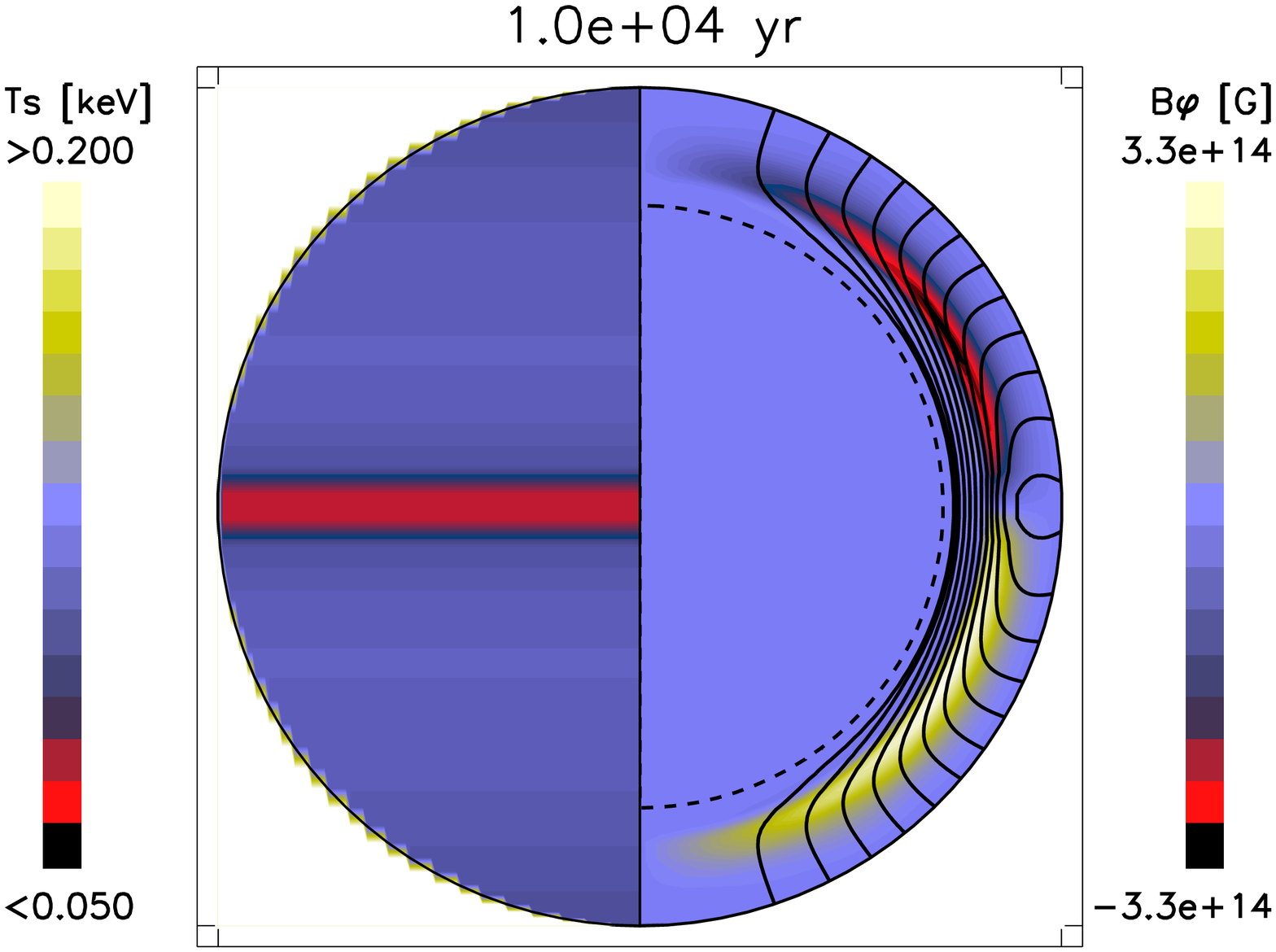}
\includegraphics[width=.32\textwidth]{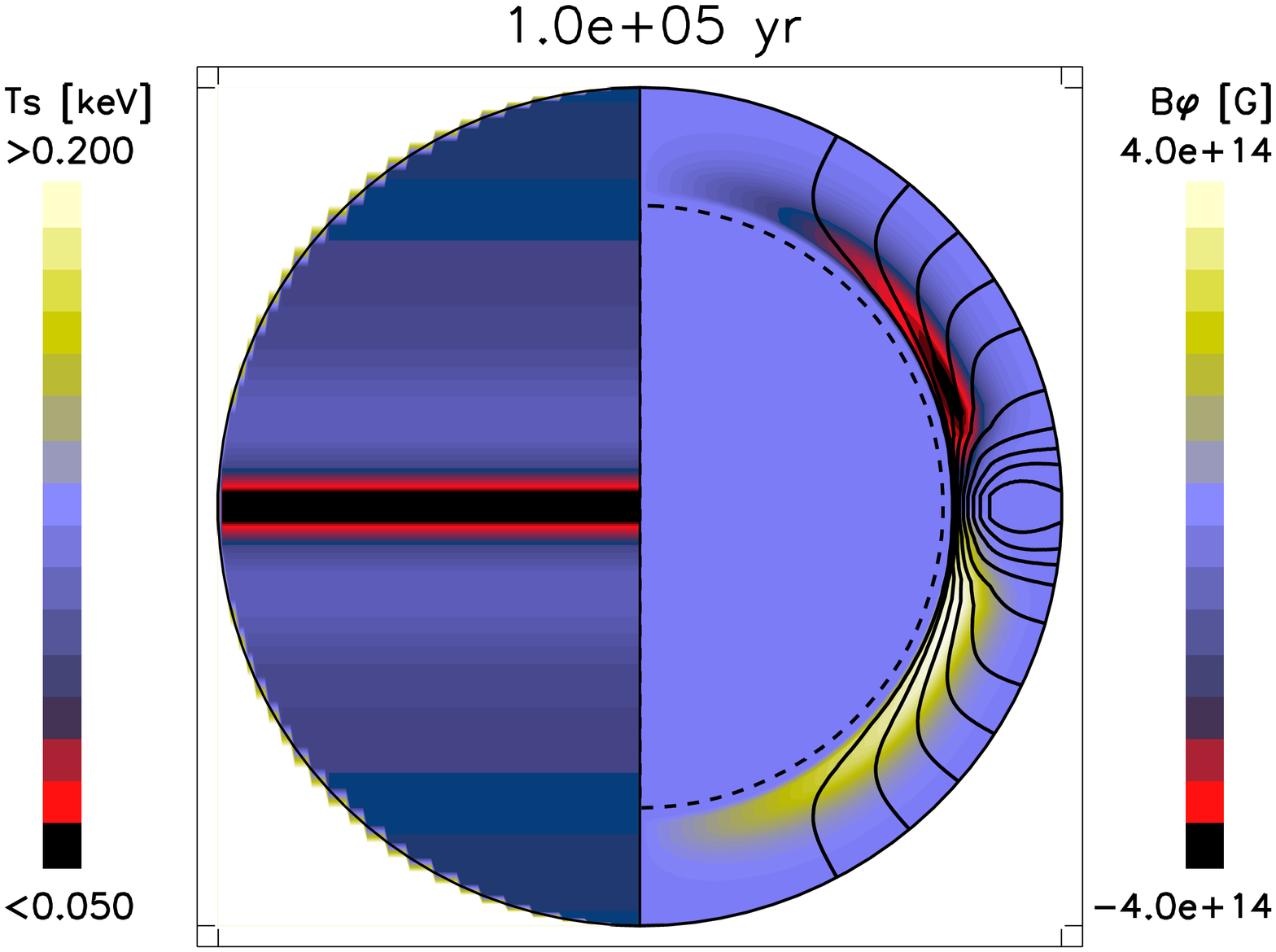}\\
\vskip0.2cm
\includegraphics[width=.32\textwidth]{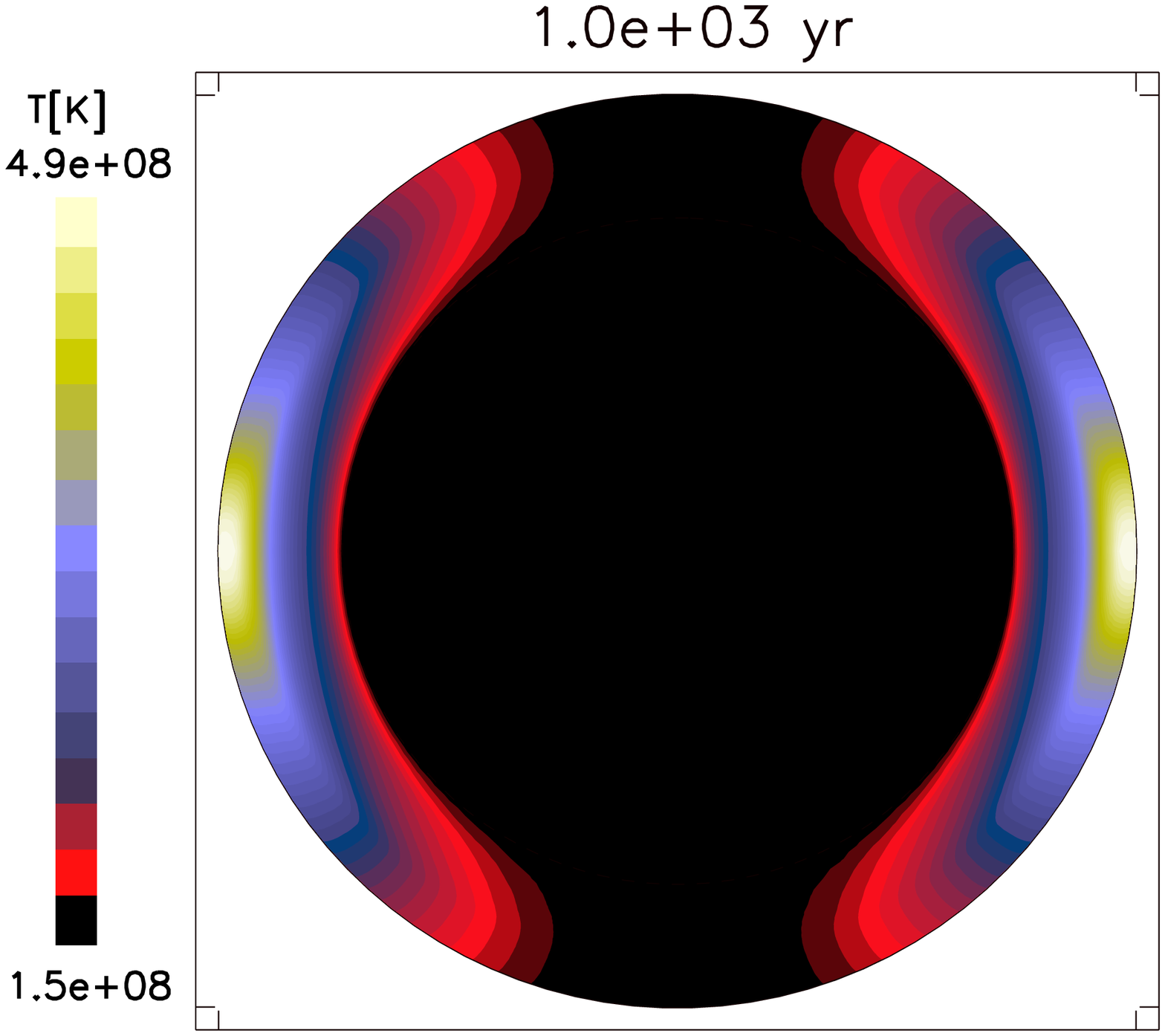}
\includegraphics[width=.32\textwidth]{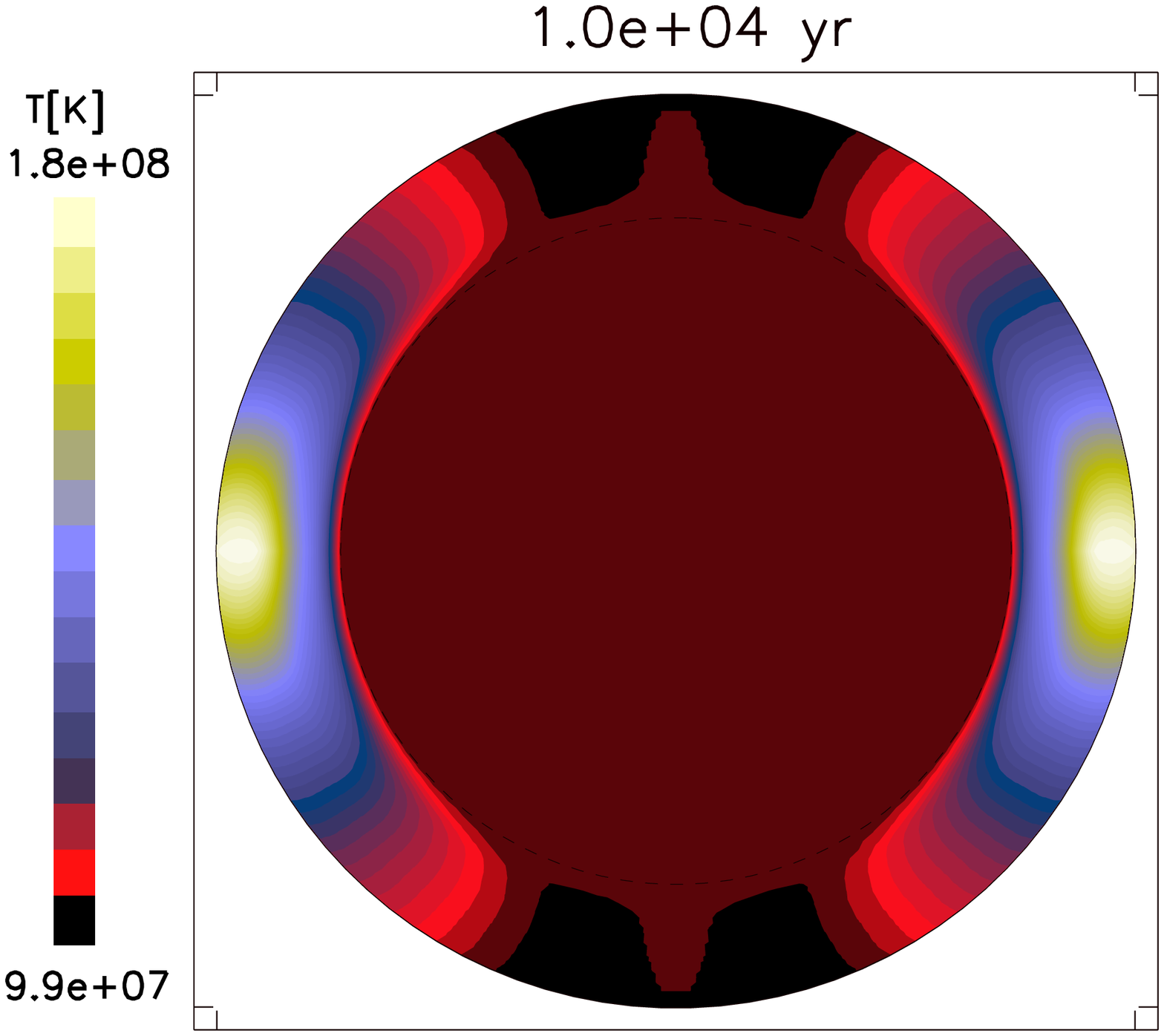}
\includegraphics[width=.32\textwidth]{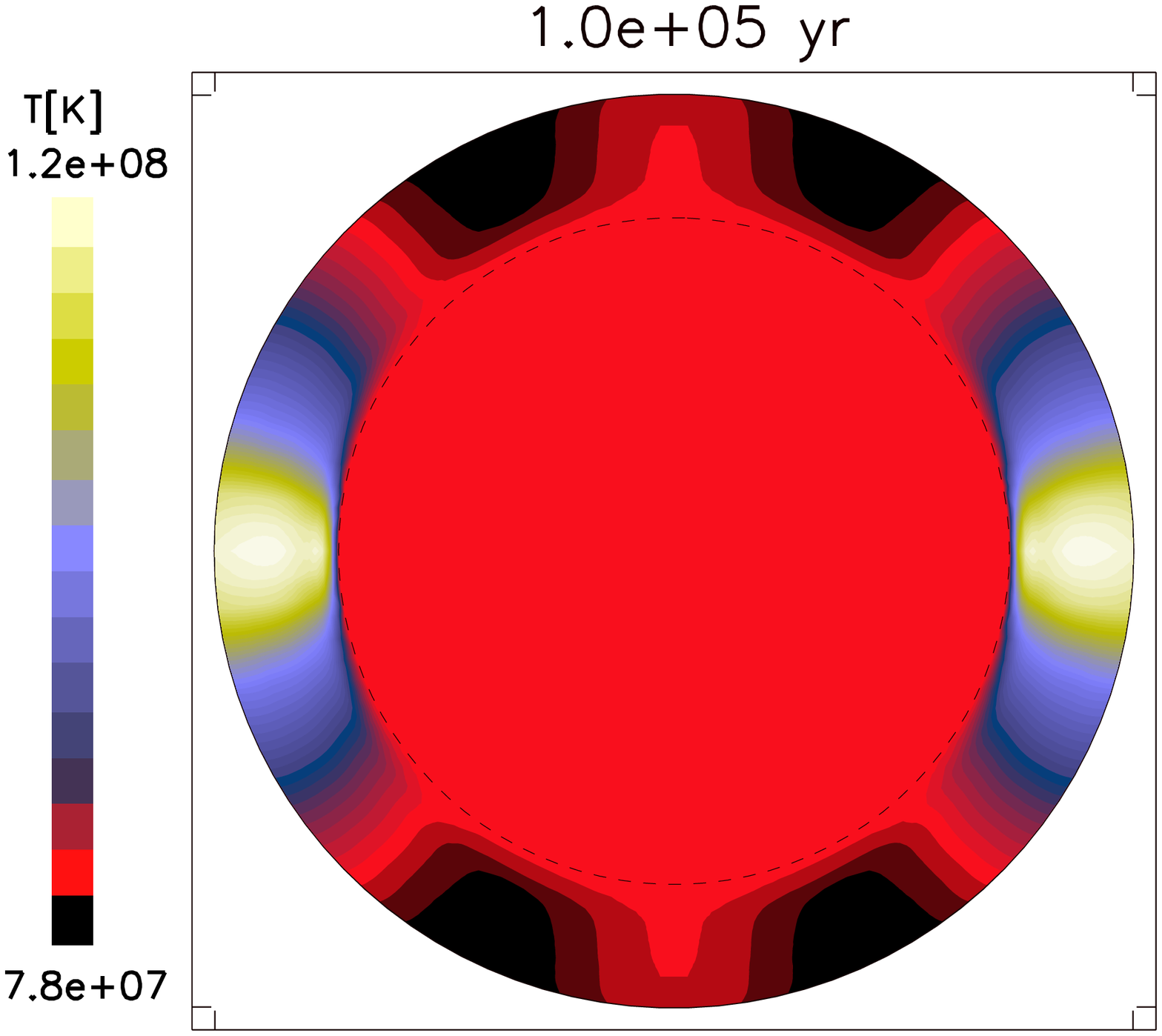}
\caption{Snapshots of the evolution of a crust-confined field model at $10^3, 10^4, 10^5$ yr, from left to right. Top panels: the left hemisphere shows in color scale the surface temperature, while the right hemisphere displays the magnetic configuration in the crust. Black lines are the poloidal magnetic field lines, while color scale indicates the toroidal magnetic field intensity. Bottom panels: temperature map inside the star. In all panels, the thickness of the crust has been enlarged by a factor of 4 for visualization purposes. Taken from \cite{viganothesis}.}
 \label{fig:b14_evo}
\end{figure}

In Fig.~\ref{fig:b14_evo} we show the evolution of the internal NS magnetic field with an initially purely dipolar component $B_p^0=10^{14}$ G, confined to the crust. After $\sim 10^3$ yr, the poloidal dipolar field has generated a mainly quadrupolar toroidal magnetic field due to the Hall effect, with a maximum strength of the same order of the poloidal magnetic field, with $B_\varphi$ being negative in the northern hemisphere and positive in the southern hemisphere. Then, the toroidal magnetic field rules the evolution, dragging the currents into the inner crust (see middle panels), and compressing the magnetic field lines. These combined effects result in fast dissipation. See also \cite{gourgouliatos14} for a detailed analysis of the Hall dynamics in the crust. 

After this phase, most of the current circulates close to the crust/core interface. Therefore, the dissipation of magnetic energy is regulated by the resistivity in this region, which is mainly controlled by the impurity parameter, if this is large enough. In this model, we set $Q_{\rm imp}=100$ close to the interface, where nuclear physics predicts the existence of the pasta phase. Such large value leads to a rapid decay of the magnetic field: this has a direct imprint on the observable rotational properties, by means of the evolution of the external dipolar component $B_p$, regulated by Eq.~(\ref{eq:ppdot_spindown}).

Ohmic dissipation and the associated heat deposition changes the map of the internal temperature. In the bottom panels of Fig.~\ref{fig:b14_evo} we show the crustal temperature evolution. At $t=10^3$ yr, the equator is hotter than the poles by a factor of 3. As the evolution proceeds and currents are dissipated, the temperature reflects the change of geometry of the poloidal magnetic field lines (see Fig.~\ref{fig:b14_evo}) and the anisotropy becomes weaker.

Strong tangential components ($B_\theta$ and $B_\varphi$) in the crust and, more importantly, in the envelope, can insulate the surface from the interior. For a dipolar geometry, the magnetic field is nearly radial at the poles, and these are thermally connected with the interior, while the equatorial region is insulated by tangential magnetic field lines. Thus, if the core is warmer than the crust, the polar regions will be warmer than the equatorial regions, unless the Ohmic dissipation heats up the equatorial regions.

\subsection{The unification of the NS zoo.}\label{sec:unification}

\begin{figure}[t]
 \centering
\includegraphics[width=.47\textwidth]{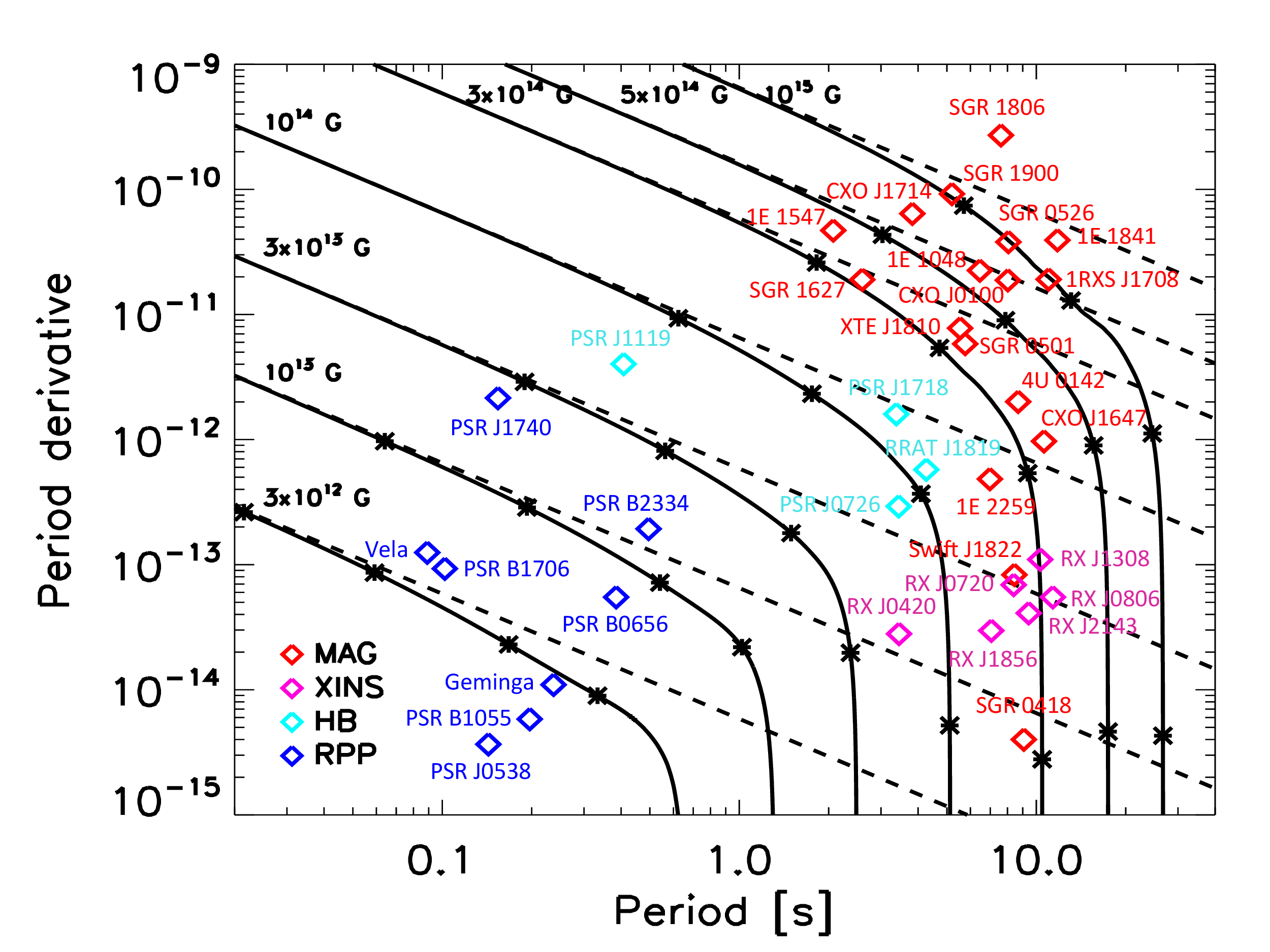}
\includegraphics[width=.47\textwidth]{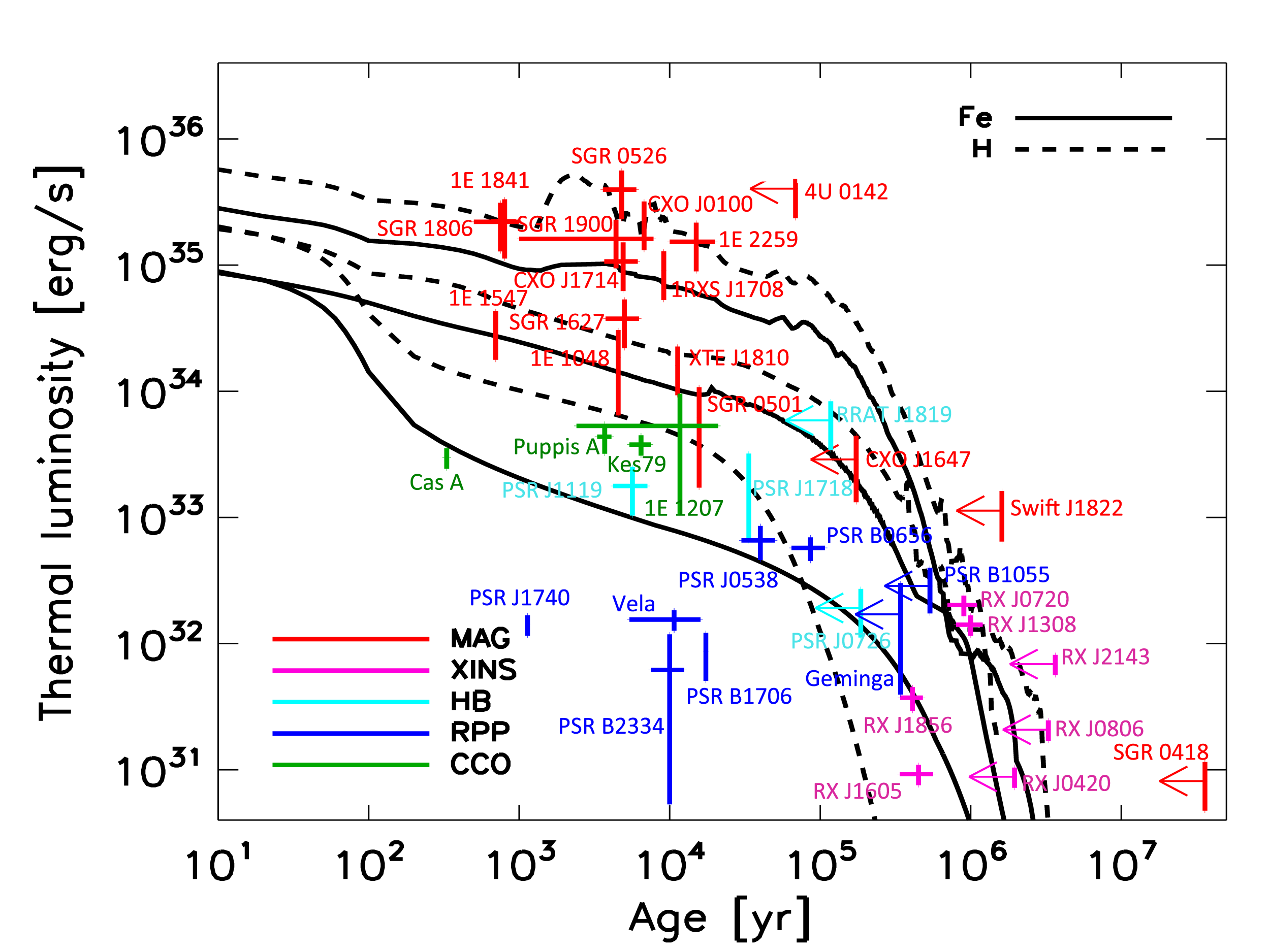}
\caption{Left: evolutionary tracks in the $P$--$\dot{P}$ diagram with $B_p^0= 3\times10^{12}, 10^{13}, 3\times10^{13}, 10^{14}, 3\times10^{14}, 10^{15}$ G. Asterisks mark the real ages $t=10^3,10^4,10^5,5\times10^5$ yr, while dashed lines show the tracks followed in absence of field decay. Right: comparison between observational data and theoretical cooling curves. Models A with $B_p^0=0, 3\times 10^{14}, 3\times 10^{15}$ G are shown for iron (solid) and light-element envelopes (dashed). Taken from \cite{viganothesis}.} 
 \label{fig:cooling_data}
\end{figure}

In the left panel of Fig.~\ref{fig:cooling_data} we show the evolutionary tracks in the $P$--$\dot{P}$ diagram for different initial values of the magnetic field strength, compared to the measured timing properties of X-ray pulsars, resulting from the evolution regulated by the evolution of $B_p(t)$. Asterisks mark different ages ($t=10^3,10^4,10^5,5\times10^5$ yr), while dashed lines show evolutionary tracks that the star would follow without magnetic field decay. $B_p$ is almost constant during an initial epoch, $t\lesssim 10^3-10^5$ yr, which depends on the initial $B_p^0$: stronger initial fields decay faster than weaker ones. Evolutionary tracks link groups of sources that belong to different classes,
for example, the magnetars with the XINSs, which in this framework are simply old, evolved magnetars.

The spin periods of the known isolated X-ray pulsars shows a sharp upper limit of $\sim 12$ s in the distribution, with most magnetars clustering in a narrow range around $\sim 10$ s. Why none of the middle age magnetars, or the older X-ray pulsars have longer periods? For X-ray pulsars, there is no reason to expect any selection effect. The likely answer is that magnetic field decays as the NS gets older \cite{colpi00}, thus its spin-down rate becomes too slow to lead to longer rotation periods during the time it is still bright enough to be visible as an X-ray pulsar. In particular, if strong currents circulate in the crust, and a high value of $Q_{imp}$ in the inner crust is employed \cite{pons13}, the vertical drop of the track in the $P$--$\dot{P}$ diagram gives a natural explanation to the observed upper limit to the rotation period of isolated X-ray pulsars, and the observed clustering of sources in the range $P=2$--$12$ seconds, while $\dot{P}$ varies over six orders of magnitude. On the other hand, in models with low impurity parameter in the whole inner crust, the magnetic field decay is not fast enough, and the period keeps increasing due to the slower dissipation of the magnetic field, which in principle predicts that pulsars of longer periods (20-100 s) should be visible. The slow release of magnetic energy through Joule heating keeps the NS bright and visible much longer than for models with large impurity in the pasta phase: magnetars with large periods should definitely be seen.

Recent population synthesis studies \cite{gullon14,gullon15} seem to favour intermediate values of $Q_{\rm imp}\sim 25$ in the innermost part of the crust, with no constrain about its value in the outer crust, which, instead, has to be inferred from the study of the thermal relaxation after type I X-ray outbursts \cite{brown09}.

\subsection{Cooling curves}

In light of these results, we can compare the data with the cooling curves predicted by the magneto-thermal evolution, for a range of magnetic field strengths. In the right panel of Fig.~\ref{fig:cooling_data} we show the luminosity as a function of time for different values of the initial magnetic field up to $B_p^0=3\times 10^{15}$ G, together with the same observational data.
Compared to the non-magnetized cooling curves, the most relevant difference is that the inclusion of the magnetic field allows us to explain objects with high luminosities. Magnetic fields above $B\gtrsim 10^{14}$ G are strong enough to noticeably heat up the crust and power the observed X-ray radiation. The cooling time-scale for strongly magnetized objects is $\sim$ one order of magnitude larger than for weakly magnetized NSs. Our results show that most magnetars, XDINSs and high-B pulsars can be reconciled with the theoretical models for a narrow range of initial magnetic fields, between $B_p^0\sim1$--$5 \times 10^{14}$ G. 

For the most luminous object that would belong to this group, 4U~0142, we do not have any alternative estimate of the age. However note that it is quite similar in both timing properties and luminosity to 1E~2259, whose kinematic age inferred for the SNR~CTB109 associated with 1E~2259 is about $10^4$ yr \cite{castro12}, more than one order of magnitude smaller than the characteristic age. We note that for these two objects it is difficult to reconcile the observed timing properties and luminosity, even with more extreme models (very strong toroidal magnetic field).

\section{Conclusions}\label{sec:conclusions}

We have summarized the main results obtained in the PhD thesis \cite{viganothesis}, which dealt with theoretical and observational aspects related to magnetic fields in NSs, which play a key role in determining the observational properties.

Our 2D magneto-thermal code includes the Joule and Hall terms throughout the full evolution. The Hall term enhances the dissipation of energy over the first $\sim 10^6$~yr of NS life, with respect to the purely resistive case. Our magneto-thermal simulations and the most recent population synthesis works \cite{gullon14,gullon15} help to paint a unified picture of the variety of observational properties of isolated NSs, and their evolutionary paths. In order to compare theoretical models with observations, we have revised the observational data for all currently known isolated NSs with clearly detected surface thermal emission. The sample includes magnetars, RPPs, XINSs, and CCOs. For all of them we have performed a homogeneous, systematic analysis of the X-ray spectra, inferring the thermal luminosity, and trying to reduce at the minimum the systematic errors due to different data analysis and modeling.

The differences in NS properties can be explained by their birth properties, in particular, the magnetic field strength. For objects with $B_p\lesssim 10^{14}$~G, the magnetic field has little effect on the luminosity. These objects, of which the RPPs are the most notable representatives, have luminosities compatible with the predictions by standard cooling models. Among them, only a few sources show evidence for enhanced cooling. On the other hand, the evolutionary models with initial magnetic fields $B_p^0\sim 3$--$5 \times 10^{14}$~G are compatible with most magnetars and their likely descendants, the XINSs. In a few extreme cases, 
we need higher initial magnetic fields, up to $B_p^0=10^{15}$~G.
Another important result is that the maximum observed spin period of isolated X-ray pulsars and the large luminosity of magnetars 
places two strong constraints. First, strong electrical currents circulating in the crust is required to explain both the timing properties 
and their high luminosities. Second, the existence of a highly resistive layer in the inner crust of NSs is required in order to explain the absence of detected pulsars with $P\gtrsim 12$ s.
Deeper observations and new discoveries of magnetars and $X$-ray pulsars will allow to further constrain the theoretical models.

Future extensions of this work includes the development of a 3D code, and a less simplified, challenging inclusion of the dynamics in the core, including: ambipolar diffusion or the interaction between superfluid vortices and superconductive flux-tubes of protons. Furthermore, more detailed population synthesis studies are needed constrain the main parameters regulating the magneto-thermal evolutionary models.

\section*{Acknowledgments}

DV thanks the SEA for the support and the prize as the best Spanish PhD thesis in Astronomy and Astrophysics 2013. The PhD research was supported by the grants AYA 2010-21097-C03-02 and a fellowship from the \textit{Prometeo} program for research groups of excellence of the Generalitat Valenciana (Prometeo/2009/103). NR is supported by an NWO Vidi Award. We also acknowledge support from the grants AYA-2013-42184-P, Prometeu II/2014/069 and from the COST action NewCompstar (MP1304). DV personally thanks the Department of Applied Physics of U. of Alicante for the great time shared.

\bibliography{biblio}

\end{document}